\documentclass[twocolumn,11pt]{article}

\newif\ifpdf
\ifx\pdfoutput\undefined
  \pdffalse
\else
  \pdfoutput=1
  \pdftrue
\fi

\ifpdf
  \usepackage[pdftex]{graphicx}
\else
  \usepackage{graphicx}
\fi

\usepackage[tbtags]{amsmath}

\usepackage{amsfonts}
\usepackage{amssymb}

\usepackage{chicago}

\renewcommand{\vec}[1]{\mbox{\boldmath $#1$}}
\newcommand{\mat}[1]{{\mbox{\bfseries \rmfamily#1}}}

\newcommand{\transp}{{\rm t}}
\newcommand{\etal}{et al.}

\begin{document}

\author{Claus O. Wilke\\\it Digital Life Laboratory\\\it
  California Institute of Technology, Mail-Code 136-93\\\it Pasadena, CA
  91125\\\it wilke@caltech.edu}
\title{Selection for Fitness vs.\ Selection for Robustness\\in RNA Secondary Structure Folding}

\maketitle

\newpage

\begin{abstract}
  We investigate the competition between two quasispecies residing on two
  disparate neutral networks. Under the assumption that the two neutral
  networks have different topologies and fitness levels, it is the mutation
  rate that determines which quasispecies will eventually be driven to
  extinction. For small mutation rates, we find that the quasispecies residing
  on the neutral network with the lower replication rate will disappear. For
  higher mutation rates, however, the faster replicating sequences may be
  outcompeted by the slower replicating ones in case the connection density on
  the second neutral network is sufficiently high. Our analytical results
  are in excellent agreement with flow-reactor simulations of replicating RNA
  sequences.
\end{abstract}

\bigskip
Keywords: quasispecies, mutant cloud, neutral networks, RNA secondary structure folding, selection of robustness

\bigskip
At high mutation rates, the number of mutated offspring generated in a
population far exceeds the number of offspring identical to their parents. As a
result, a stable cloud of mutants, a so-called
quasispecies~\shortcite{EigenSchuster79,%
  Eigenetal88,Eigenetal89,Nowak92,Wilkeetal2001a}, forms around the fastest
replicating genotypes. Experimental evidence in favor of such a persistent
cloud of mutants is available from RNA
viruses~\shortcite{Steinhaueretal89,DomingoHolland97,BurchChao2000} and in
vitro RNA replication~\shortcite{Biebricher87,BiebricherGardiner97}; both are
cases in which a high substitution rate per nucleotide is
common~\shortcite{Drake93}.  The existence of a quasispecies has important
implications for the way in which selection acts, because the evolutionary
success of individual sequences depends on the overall growth rate of the
quasispecies they belong to. As a consequence, organisms with a high
replication rate that produce a large number of offspring with poor fitness
can be outcompeted by organisms of smaller fitness that produce a larger
number of also-fit offspring~\shortcite{SchusterSwetina88}.  Similarly, if a
percentage of the possible mutations is neutral, and the majority of the
non-neutral mutations is strongly deleterious, then the growth rate of a
quasispecies depends significantly on the connection density (the number of
nearby neutral mutants of an average viable genotype) of the neutral
genotypes~\shortcite{vanNimwegenetal99b}.  Therefore, a neutral network (a set
of closely related mutants with identical fitness) with high connectivity can
be advantageous over one with higher fitness, but lower connectivity. Here, we
are interested in this latter possibility. In particular, we investigate the
competition of two quasispecies residing on separate neutral networks with
different connection densities and replication rates, and determine under what
conditions selection favors the more fit (i.e., of higher replication rate) or
the more robust (more densely connected) mutant cloud. Our approach is closely
related to the study of holey landscapes, in which all genotypes are
classified into either viable or inviable
ones~\shortcite{GavriletsGravner97,Gavrilets97}. However, we extend this
picture by further subdividing the viable genotypes into two groups
with different replication rates.

The paper is organized as follows. First, we describe a simple model of a
quasispecies on a single neutral network, and demonstrate that the model is
consistent with simulations of RNA sequences. Then, based on this model, we
present a model of two competing quasispecies, and compare the second model
with simulation results as well. Following that, we study the probability of
fixation of a single advantageous mutant that arises in a fully formed
quasispecies.  Finally, we discuss the implications of our results and give
conclusions.

\begin{center}
\sc Population Dynamics on a Single Neutral Network
\end{center}

Before we can address the competition of two quasispecies, we need a good
description of a single quasispecies on a neutral network. A fundamental
contribution to this problem has been made by \shortciteN{vanNimwegenetal99b},
who showed that the average fitness of a population on a neutral network is
determined only by the fitness of the neutral genotypes, the mutation rate,
and the largest eigenvalue of the neutral genotypes' connection matrix. The
connection matrix is a symmetric matrix with one row/column per neutral
genotype. It holds a one in those positions where the row- and the
column-genotype are exactly one point-mutation apart, and a zero otherwise. In
theory, the formalism of van Nimwegen \etal\ describes a population on a
neutral network well. However, the exact connection matrix is normally not
known, which implies that we cannot calculate the population dynamics from
first principles. Nevertheless, we can base a very simple model on the fact --
also established by van Nimwegen \etal -- that the average neutrality in the
population, which is exactly the largest eigenvalue of the connection matrix,
is independent of the mutation rate. The main assumption of our simple model
is that the population behaves as if all sequences in the population had the
same neutrality $\nu$, where $\nu$ is given by the average neutrality in the
population. Moreover, we consider genetic sequences of length $l$, and assume
a per-symbol copy fidelity of $q$.  Then, the effective copy fidelity or
neutral fidelity~\shortcite{OfriaAdami2001} $Q$, i.e., the probability with
which on average a viable sequence gives birth to offspring that also resides
on the neutral network, is given by
\begin{align}\label{eq:def-Q}
  Q &= [1-(1-q)(1-\nu)]^l\notag\\
   & \approx e^{-l(1-q)(1-\nu)}\,.
\end{align}
Now, we can devise a two-concentration model in which $x_1(t)$ is the total
concentration of all sequences on the neutral network, and $x_d(t)$ is the
concentration of sequences off the network (these sequences are assumed to
replicate so slowly that their offspring can be neglected). The two
quantities satisfy the equations
\begin{subequations}
\begin{align}\label{eq:two-conc-1}
\dot x_1(t) &= w_1 Q x_1(t) - e(t)x_1(t)\,,\notag\\
\dot x_d(t) &= w_1 (1-Q) x_1(t) - e(t)x_d(t)\,,
\end{align}
\end{subequations}
where $w_1$ is the fitness of the sequences on the neutral network, and $e(t)$ is the excess production (or mean fitness in the population)
$e(t)=w_1x_1(t)$. Equation~\eqref{eq:two-conc-1} can be integrated
directly. We find
\begin{equation}
  x_1(t) = \frac{Q x_1(0)}{x_1(0) + [Q-x_1(0)]e^{-w_1Qt}}\,.
\end{equation}
In the steady state ($t\rightarrow\infty$), this implies that the
concentration of sequences on the network is equal to the effective fidelity
$Q$,
\begin{equation}\label{eq:steady-state-conc}
  x_1 = Q = e^{-l(1-q)(1-\nu)}\,.
\end{equation}
Therefore, by measuring the decay of the concentration of sequences on the
neutral network as a function of the copy fidelity $q$, we can estimate the
population neutrality $\nu$.

Note that the above description of the evolving population is similar to
the one presented by~\shortciteN{Reidysetal2001}, with one important
conceptual difference. The article by \shortciteN{Reidysetal2001} was
completed before van Nimwegen \etal's work was available, and therefore it was
not clear what their effective fidelity did actually relate to. Here, on the
other hand, we know that $Q$ depends only on the copy fidelity per nucleotide,
$q$, and the average population neutrality $\nu$, which is independent of $q$
and could be calculated exactly if the connection matrix of the neutral
genotypes was known.

We have measured the average equilibrium concentration $x_1$ of sequences on
the network for RNA secondary structure folding. RNA folding is a reliable
test case, and has been applied to a wide array of different questions related
to the dynamics of
evolution~\shortcite{Fontanaetal93,Huynenetal96,FontanaSchuster98,SchusterFontana99,AncelFontana2000,Reidysetal2001}.
We simulated a flow reactor using the Gillespie
algorithm~\shortcite{Gillespie76}, and performed the RNA folding with the
Vienna package~\shortcite{Hofackeretal94}, version 1.3.1, which uses the
parameters given by~\shortciteN{Walteretal94}. The carrying capacity was set
to $N=1000$ sequences, and the reactor was initially filled with 1000
identical copies of a sequence that folded into a given target structure.
Sequences folding into the target structure were replicating with rate one per
unit time, and all other sequences with rate $10^{-6}$ per unit time. We let
the reactor equilibrate for 50 time steps, and then measured the average
concentration of correctly folding sequences over the next 150 time steps.

Results for the two different target structures depicted in
Fig.~\ref{fig:folds} are shown in Fig.~\ref{fig:decay}. In both cases, we see
a very clear exponential decay. Up to a mutation rate of $0.05$, which is
quite high for the sequences of length $l=62$ we are considering here, we
cannot make out a significant deviation from a straight line in the log-linear
plot. This verifies the applicability of our simple model to evolving RNA
sequences. Note that our simulations also show a significant difference in the
effective neutrality of the two structures, which will be of importance in the
next section.

\begin{center}
\sc Two Competing Quasispecies
\end{center}

\begin{center}
\it Analytical Model
\end{center}

Above, we have established a simple description for a quasispecies residing on
a single neutral network. In a similar fashion, we can treat the competition
of two quasispecies residing on separate networks. We classify all sequences
into three different groups: sequences on network one, sequences on network
two, and dead sequences (sequences that replicate much slower than sequences
on either of the two networks, or do not replicate at all). We denote the
respective relative concentrations by $x_1$, $x_2$, and $x_d$. We make the
further assumption that all sequences within a neutral network $i$ have the
same probability $Q_i$ to mutate into another sequence on network $i$, and we
neglect mutations from one network to the other. The probability to
fall off of a network $i$ is hence $1-Q_i$. The differential equations for an
infinite population are then:
\begin{subequations}\label{eq:basic-model}
\begin{align}
  \dot x_1(t) &= w_1 Q_1 x_1(t) - e(t) x_1(t)\,,\\
  \dot x_2(t) &= w_2 Q_2 x_2(t) - e(t) x_2(t)\,,\\
  \dot x_d(t) &= w_1 (1-Q_1) x_1(t) + w_2 (1-Q_2) x_2(t) - e(t) x_d(t)\,,
\end{align}
\end{subequations}
where $w_1$ and $w_2$ are the fitnesses of sequences on network one or two,
respectively, and $e(t)$ is the excess production $e(t)=w_1 x_1(t) + w_2
x_2(t)$. In order to solve Eq.~\eqref{eq:basic-model}, it is useful to
introduce the matrix
\begin{equation}
  \mat W = \begin{pmatrix} w_1 Q_1 & 0 & 0 \\
                           0 & w_2 Q_2 & 0 \\
                           w_1(1-Q_1) & w_2(1-Q_2) & 0 \end{pmatrix}\,.
\end{equation}
We further need the exponential of $\mat W$, which is given by
\begin{equation}
  \exp(\mat W t) = \begin{pmatrix}
    e^{w_1Q_1t} &0 & 0\\
    0 & e^{w_2Q_2t} &0\\
    \frac{1-Q_1}{Q_1}(e^{w_1Q_1 t} - 1) &  \frac{1-Q_2}{Q_2}(e^{w_2Q_2 t} - 1)
 & 1 \end{pmatrix}
\end{equation}
Now, if we combine the concentrations $x_1$, $x_2$, $x_d$ into a vector
$\vec x = (x_1, x_2, x_d)^\transp$, we find
\begin{equation}\label{eq:formal-sol}
  \vec x(t) = \exp(\mat W t)\cdot \vec x(t) /[\hat{\vec e}\cdot \exp(\mat W t)
  \cdot \vec x(0)]
\end{equation}
with $\hat{\vec e}:=(1,1,1)$. The denominator on the right-hand side of
Eq.~\eqref{eq:formal-sol} corresponds to the cumulative excess production
$e_{\rm cum}(t) = \int_0^t e(t)\, dt$, which is given by
\begin{align}
  e_{\rm cum}(t) &= \hat{\vec e}\cdot \exp(\mat W t) \cdot \vec x(0)\notag\\
 &= \frac{x_1(0)}{Q_1}(e^{w_1Q_1 t}+Q_1 - 1)\notag\\
 &\quad + \frac{x_2(0)}{Q_2}(e^{w_2Q_2
 t}+Q_2 - 1) + x_d(0)\,.
\end{align}
The solution to Eq.~\eqref{eq:basic-model} follows now as
\begin{subequations}\label{eq:3conc-sol}
\begin{align}\label{eq:3conc-sol1}
  x_1(t) &= \frac{e^{w_1 Q_1 t}}{e_{\rm cum}(t)} x_1(0)\,,\\
\label{eq:3conc-sol2}
  x_2(t) &= \frac{e^{w_2 Q_2 t}}{e_{\rm cum}(t)} x_2(0)\,,\\
  x_d(t) &= \frac{1}{Q_1Q_2e_{\rm cum}(t)}\Big[(e^{w_1Q_1t}-1)(1-Q_1) Q_2 x_1(0) \notag\\
 & \qquad + (e^{w_2Q_2t}-1)(1-Q_2) Q_1 x_2(0) + Q_1 Q_2 x_d(0)\Big]\,.
\end{align}
\end{subequations}
There exist two possible steady states. If $w_1Q_1>w_2Q_2$, then for
$t\rightarrow\infty$ we have $x_1=Q_1$, $x_2=0$, $x_d=1-Q_1$. If
$w_1Q_1<w_2Q_2$, on the other hand, the steady state distribution if given by
$x_1=0$, $x_2=Q_2$, $x_d=1-Q_2$. The most interesting situation occurs when
for a given $w_1$ and $w_2$, the steady state depends on the mutation rate.
This happens if $w_1>w_2$, but $\nu_1<\nu_2$, or vice versa. Namely, if we
express $Q_i$ as given in Eq.~\eqref{eq:def-Q}, we obtain from $w_1Q_1=w_2Q_2$
the critical copy fidelity
\begin{equation}\label{eq:q-crit}
  q_{\rm c} = 1-\frac{\ln(w_2/w_1)}{l(\nu_1-\nu_2)}\,.
\end{equation}
Clearly, $q_{\rm c}$ can only be smaller than one if either $w_1>w_2$ and
$\nu_1<\nu_2$ or vice versa. Therefore, this is a necessary (though not
sufficient) condition for the existence of two qualitatively different steady
states in different mutational regimes.  In the language of physics, the
transition from one of the two steady state to the other is a first order
phase transition~\shortcite{Stanley71}. The transition is of first order
because the order parameter (which we can define to be either $x_1$ or $x_2$)
undergoes a discontinuous jump from a finite value to zero at the critical
mutation rate.

The two phases are not just a mathematical curiosity, they have important
biological interpretations. The phase in which the sequences with the larger
$w_i$ survive can be considered the ``normal'' selection regime, i.e.,
selection which favors faster replicating individuals. We will refer to this
situation as the phase of ``selection for replication speed''. In the other
phase, however, the situation is exactly reversed, and the sequences with the
lower intrinsic replication rate $w$ prevail. In this phase, the amount of
neutrality (or the robustness against mutations) is more important, and we
will consequently refer to this situation as the phase of ``selection for
robustness''. In Fig.~\ref{fig:phases}, we show two example phase
diagrams. These diagrams demonstrate that the selection for robustness is not
a pathological situation occurring only for extremely rare sets of parameters,
but that in fact both phases have to be considered on equal grounds, none of
them can be singled out as the more common one. In particular, as the ratio
between $w_1$ and $w_2$ approaches unity, the selection for robustness becomes
more and more important.

\begin{center}
\it Simulation Results
\end{center}

As in the case of a single quasispecies on a neutral network, we have tested
our predictions with simulations of self-replicating RNA sequences in a flow
reactor. We assumed that sequences folding into either Fold~1 or~2
(Fig.~\ref{fig:folds}) were replicating with rates $w_1=1$ and $w_2=1.1$,
respectively, while all other folds had a vanishing replication rate. In all
results presented below, we initialized the flow reactor with 50\% of the
sequences folding into Fold~1, and the remaining sequences folding into
Fold~2.

Figure~\ref{fig:comp0_01} shows a comparison between Eq.~\eqref{eq:3conc-sol}
and four example runs. Apart from finite size fluctuations, which are to be
expected in a simulation with $N=1000$, the analytic expression predicts the
actual population dynamics well.

In Fig.~\ref{fig:fave-vs-mutrate}, we present measurements of the
concentrations $x_1(t)$ and $x_2(t)$ as functions of the mutation rate $1-q$,
for a fixed time $t=200$. The points represent results averaged over 25
independent simulations, and the lines stem from Eq.~\eqref{eq:3conc-sol}. In
agreement with the predictions from our model, we observe two selection
regimes, one in which the faster replicating sequences dominate, and one in
which the sequences with the higher neutrality have a selective advantage. The
transition between the two phases occurs in this particular case approximately
at $q=0.98$, and both the analytical model and the simulations agree well on
this value. As is typical for a phase transition, the fluctuations close to
the transition point increase significantly, and the time until either of the
two quasispecies has gone extinct diverges (the latter point can be seen from
the fact that close to the transition point, the disadvantageous fold is still
present in a sizeable amount, while further away it has already vanished
completely from the population).

Figure~\ref{fig:fave-vs-mutrate} also shows that for very small
populations, the predictive value of the differential equation approach
diminishes, presumably because the choice of a single effective copy fidelity
$Q$ is not justified anymore once a minimum population size has been reached.
However, as long as we are dealing with population sizes of several hundreds
or more, our analytical calculations predict the simulation results very well.


\begin{center}
\it Probability of Fixation
\end{center}

In the previous subsection, we have established that selection acts on the
product of replication rate $w$ and fidelity $Q$, rather than on the
replication rate alone. In particular, for an appropriate choice of
parameters, sequences with a lower replication rate can outcompete those with
a higher replication rate. However, the competition experiments that we
conducted in the previous section were unrealistic in so far that we assumed
equal initial concentrations of the two competing types of sequences. A more
realistic assumption is that one type (the one with the lower product
$w_iQ_i$) dominates the population, while the second type is initially
represented through only a single individual. The idea behind this scenario is
of course that the second type (with higher product $w_iQ_i$) has arisen
through a rare mutation. The question in this context is whether the second
type will be able to dominate the population, i.e., whether it will become
fixated.

In a standard population genetics scenario, the answer to the above question
is simple. If two sequences replicate with $w_1$ and $w_2$, respectively, and
mutations between the two sequences can be neglected, then a single sequence of
type 2 ($w_2>w_1$) will become fixated in a background of sequences of type 1
with probability $\pi=1-e^{-2s}\approx 2s$, where $s=w_2/w_1-1$ is the
selective advantage of the newly introduced sequence
type~\shortcite{Haldane27,Kimura64,Ewens79}.  Note, however, that this
celebrated result is only correct for a generational model with discrete time
steps. In a continuous time model, the equivalent result reads $\pi=s/(1+s)$.
This formula follows from the solution to the problem of the Gambler's
Ruin~\shortcite{Feller68,LenskiLevin85} when taking the limit of a large
population size.

Here, we are not dealing with individual sequences replicating with rate
$w_i$, but rather with quasispecies that grow with rate $w_iQ_i$. A naive way
to calculate the fixation probability in this case is simply to replace $w_i$
with $w_iQ_i$ in the expression for the selective advantage, and hope that the
result is correct. However, it is not clear from the outset that this
approach will work, because the factor $Q_i$ depends on the assumption that a
fully developed quasispecies with the appropriate mean neutrality is already
present. A single sequence struggling for fixation does not satisfy this
condition. Therefore, the actual fixation probability might deviate from the
one thus calculated, in particular in circumstances in which
a sequence with smaller replication rate is supposed to overtake an
established quasispecies of sequences with higher replication rate.

We performed fixation experiments in both the ``selection for replication
speed'' and the ``selection for robustness'' phase, in order to clarify
whether the naive approach works. In both phases, we allowed a population of
size $N=1000$ to equilibrate, and then introduced a single sequence of the
supposedly advantageous type. After 500 time steps, we determined whether the
advantageous type had vanished from the population or grown to a significant
proportion. By repeating this procedure 100 times, we obtained an estimate for
the probability of fixation. As in the previous section, we used $w_1=1$ and
$w_2=1.1$.

In Fig.~\ref{fig:fixation-prob}, we compare our simulation results to the
predicted fixation probability $\pi=s/(1+s)$. Within the accuracy of our
results, both agree well. This is particularly interesting for mutation rates
above $0.02$, where we introduce a sequence of lower replication rate into a
background of faster replicating sequences. The increased neutrality of the
introduced sequence is sufficient to let it rise to fixation in a significant
proportion of cases. Moreover, the product $w_iQ_i$ is the sole determinant of
the fixation probability. Whether the value of the product $w_iQ_i$ comes
mainly from the intrinsic growth rate $w_i$ of the sequences or from the
effective fidelity $Q_i$ does not have an observable influence on the
dynamics.

\begin{center}
\sc Discussion
\end{center}

The good agreement between our analytical model and our simulation results
demonstrates that RNA sequences evolving on a neutral network of identical
secondary structure folds are well described by only two parameters, their
intrinsic replication rate $w$ and their effective copy fidelity $Q$. In the
particular context of two competing distinct folds, we find furthermore that
only the product of $w$ and $Q$ is of importance. Indeed, it follows from
Eq.~\eqref{eq:3conc-sol} that the ratio between $x_1(t)$ and $x_2(t)$
depends only on the respective products of $w$ and $Q$,
but not on the individual values themselves.

Unlike the intrinsic replication rate $w$, which is a property of the
individual, the effective fidelity $Q$ is a group property, as it is given by
the average over all sequences in the population of the probability not to
``fall off'' the neutral network. Thus, in the regime in which $Q$ dominates
the evolutionary dynamics (the phase of selection for robustness in
Fig.~\ref{fig:phases}), the evolutionary success of an individual sequence
depends strongly on the properties of the group it belongs to. In other words,
we find that selection acts on the whole group of mutants, rather than on
individuals, despite the absence of standard factors supporting group
selection such as spatial subdivision of the population~\shortcite{Wilson79},
altruistic behavior, parental care~\shortcite{MaynardSmith93}, or mutual
catalytic support~\shortcite{Alvesetal2001}. Here, a sequence with a
comparatively high neutrality embedded into a neutral network with a poor
overall connection density will be at a disadvantage with respect to a
sequence with a comparatively low neutrality that is, however, part of a
neutral network with high connection density. The overall higher fidelity of a
population on the second network results in a larger fraction of sequences
that actually reside on the network, which in turn increases the chance that a
particular sequence will be generated as mutant offspring from some other
sequence. \shortciteN{Moyaetal2000} noted that this type of group
selection should follow from the quasispecies equations, and that populations
under this type of selection would be best described by an effective group
replication rate $r$. In the present work, we have shown that this is indeed
the case, and we can also derive $r$ (which is simply $r=wQ$) from the
quasispecies equations.  Namely, the fact that the population neutrality $\nu$
(which determines $Q$) is given by the largest eigenvalue of the connection
matrix of neutral genotypes is a direct consequence of the quasispecies
equations~\shortcite{vanNimwegenetal99b}.

\shortciteN{SchusterSwetina88} were the first to point out
that at high mutation rates, the quasispecies around the highest peak in the
landscape can disappear. They focused on situations in which the highest and
the second-highest peak in a landscape were of almost equal height, while the
immediate mutational neighborhood of the second peak was less deleterious than
the one of the first peak. As a consequence, their results seemed to imply
that the phase of 'selection of robustness' was only important in the case of
very similar peaks. Our results, on the other hand, show that
the difference in peak hight can be dramatic, if balanced by an equally
dramatic difference in robustness.

While our analytical results apply strictly speaking only to infinite
populations, we have seen that in simulations for population sizes as small as
$N=500$, the differential equation approach works well. Moreover, in our
experiments on the probability of fixation, we have seen that even very small
numbers of the advantageous group (in the extreme only a single sequence) can
rise to fixation, despite their intrinsic replication rate being smaller than
that of the currently dominating group. This result seems somewhat unintuitive
at first, but can be easily understood. The most important aspect of every
fixation event is the very first replication of the new genotype, and the
smaller its selective advantage, the more likely it is not to replicate even
once. Now, if a new mutant with a poor replication rate $w_{\rm new}$ but high
effective fidelity $Q_{\rm new}$ arises in a population that is dominated by
sequences with large intrinsic replication rate, we would intuitively assume
that the mutant will hardly ever replicate even once, and therefore will never
get a chance to employ its superior fidelity. However, this is not correct if
the effective fidelity of the dominating sequences, $Q_{\rm dom}$, is low.
From Eq.~\eqref{eq:steady-state-conc}, we find that the concentration of
sequences that actually replicate is given by $Q_{\rm dom}$. Therefore, even
though the sequences that replicate do so at a high rate, the actual number of
births that occur is small, comparable to the one in a population in which all
individuals reproduce with rate $w_{\rm dom}Q_{\rm dom}$. Therefore, the newly
introduced genotype is relatively safe from being washed out prematurely, and
fixation takes place at the predicted rate.

\begin{center}
\sc Conclusions
\end{center}

We have demonstrated that for a population in a landscape where neutral
mutants abound, the product of intrinsic replication rate $w$ and effective
copy-fidelity $Q$ is being maximized under selection, rather than the
intrinsic replication rate alone. This observation has led to the natural
distinction between two modes of selection, one in which intrinsic replication
rate is favored, and one in which robustness (high $Q$) is more important. In
the latter phase, the success of a single sequence depends strongly on the
mutant cloud the sequence belongs to. Our results thus demonstrate
that the unit of selection in molecular evolution is indeed the quasispecies,
as proposed by \shortciteN{EigenSchuster79}, and not the
individual replicating sequence. In particular, the probability of fixation of
a single advantageous mutant in an established quasispecies can be predicted
accurately with results from standard population genetics, provided we
consider the overall growth rates of the established quasispecies and the
quasispecies potentially formed by the mutant, rather than the replication
rates of mutant and established wild type.

\section*{Acknowledgments}

This work was supported by the NSF under contract No DEB-9981397. C.O.W. would
like to thank (in alphabetical order) C.~Adami for many useful comments and
suggestions; P.~Campos for double-checking fixation probabilities; W.~Fontana
for providing the original flow-reactor code; J.~Wang for writing an early
Mathematica script used in this study.

\begin{figure}[p]
\caption{\label{fig:folds}The two different folds used in this study. Both
  consist of the same number of base pairs ($l=62$), but Fold~1 has a higher
  neutrality ($\nu=0.442$) than Fold~2 ($\nu=0.366$). See also
  Fig.~\ref{fig:decay}.}
\end{figure}

\begin{figure}[p]
\caption{\label{fig:decay}Decay of the steady state concentration $x_1$ as a
  function of $1-q$ for two example secondary structures. The solid and the
  dashed line are given by $\exp[-l(1-q)(1-\nu_i)]$ with $l=62$. The values
  for $\nu_1$ and $\nu_2$ have been obtained from a fit of this expression to
  the measured data (shown as points with bars indicating the standard error).}
\end{figure}

\begin{figure}[p]
\caption{\label{fig:phases}Typical phase diagram following from
  Eq.~\ref{eq:q-crit}. We used $l=100$, $w_2=1$, and $\nu_1=0.5$, as well as
  $\nu_2=0.6$ in graph a) and $w_1=1.5$ in graph b).}
\end{figure}

\begin{figure}[p]
\caption{\label{fig:comp0_01}Concentrations $x_1(t)$ and $x_2(t)$ as functions of
  the time $t$ for a copy fidelity of $q=0.99$. The thick lines represent the
  analytic predictions from Eqs.~\eqref{eq:3conc-sol1}
  and~\eqref{eq:3conc-sol2}, and the thin lines stem from simulations with
  $N=1000$. }
\end{figure}

\begin{figure}[p]
\caption{\label{fig:fave-vs-mutrate}Concentrations $x_1(200)$ (dashed lines)
  and $x_2(200)$ (solid lines)  as
  functions of the per-nucleotide mutation rate $1-q$. The lines
  represent the analytic predictions. The points represent the average over 25
  independend simulation runs each, with bars indicating the standard
  error. We performed the simulations
  with four different population sizes, $N=5000$ (a), $N=1000$ (b), $N=500$
  (c), and $N=100$ (d). The initial concentrations in all simulations were
  $x_1(0)=x_2(0)=0.5$, $x_d(0)=0$.}
\end{figure}

\begin{figure}[p]
\caption{\label{fig:fixation-prob}Probability of fixation as a function of the
  mutation rate. Below $1-q=0.02$, we are looking at the probability of
  fixation of a single sequence of type 2 in a full population of sequences of
  type 1. Above $1-q=0.02$, we are considering the reversed configuration. The
  solid and dashed line represent the analytical prediction $\pi=s/(1+s)$, the
  points stem from simulations (bars indicate standard error).}
\end{figure}

\onecolumn

\newpage
\noindent Figure 1:

\centerline{
\includegraphics[width=16cm,angle=90]{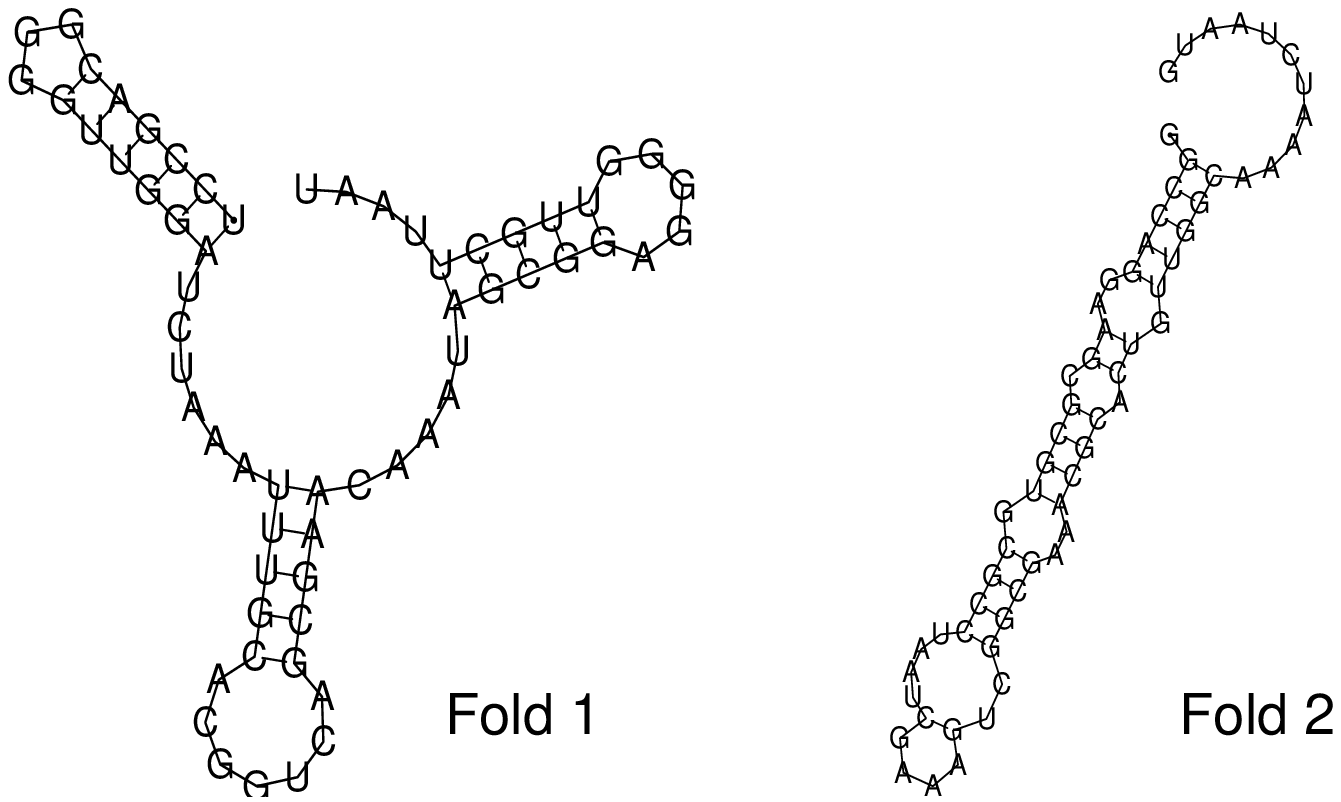}
}

\newpage
\noindent Figure 2:

\centerline{
\includegraphics[width=16cm,angle=90]{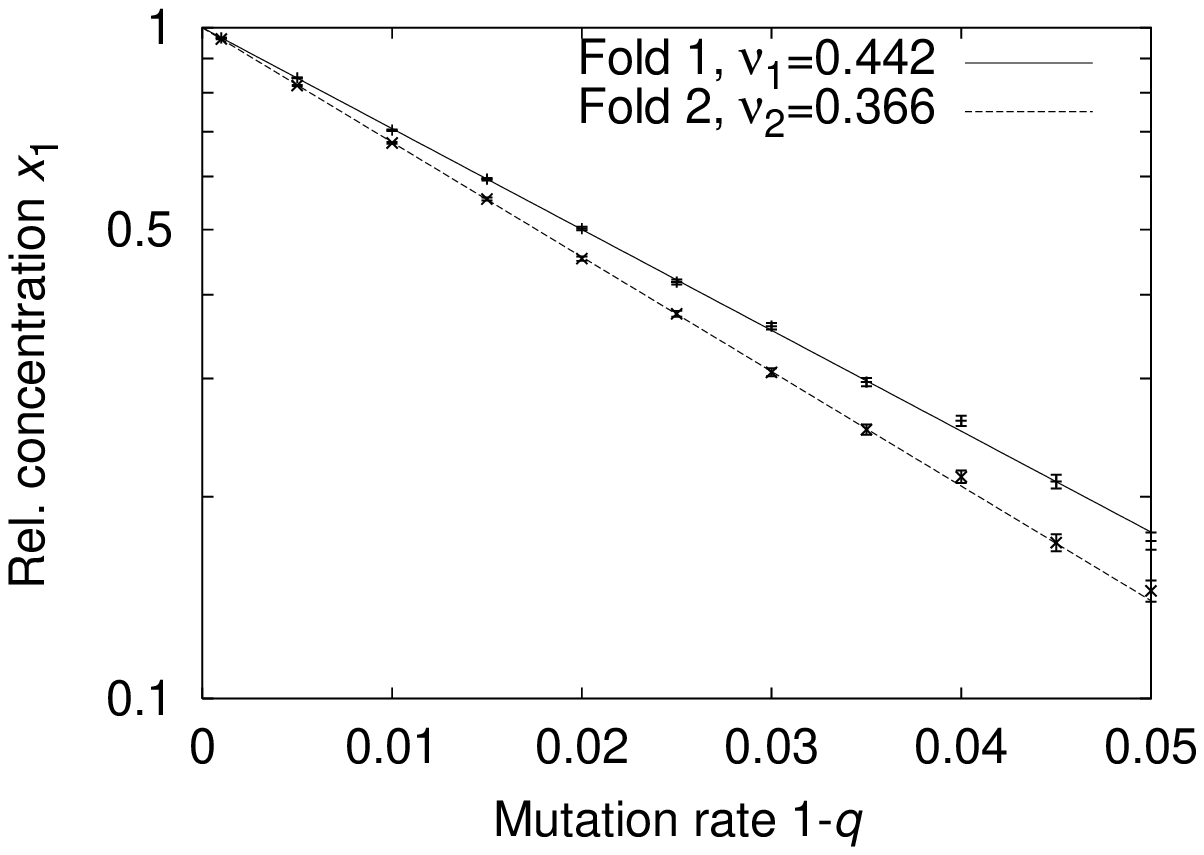}
}

\newpage
\noindent Figure 3:

\centerline{
\includegraphics[width=17cm,angle=90]{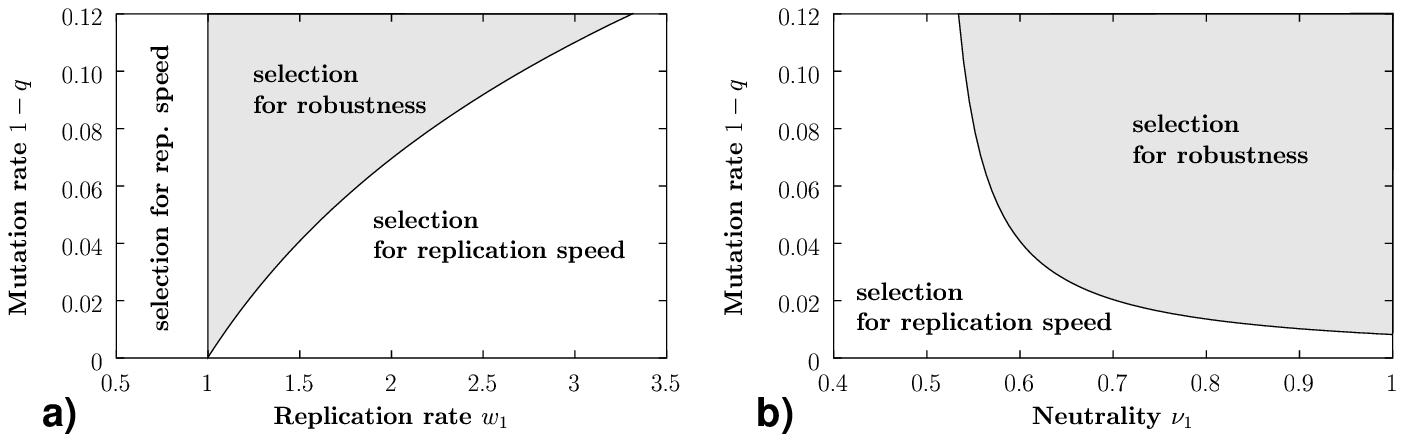}
}

\newpage
\noindent Figure 4:

\centerline{
\includegraphics[width=16cm,angle=90]{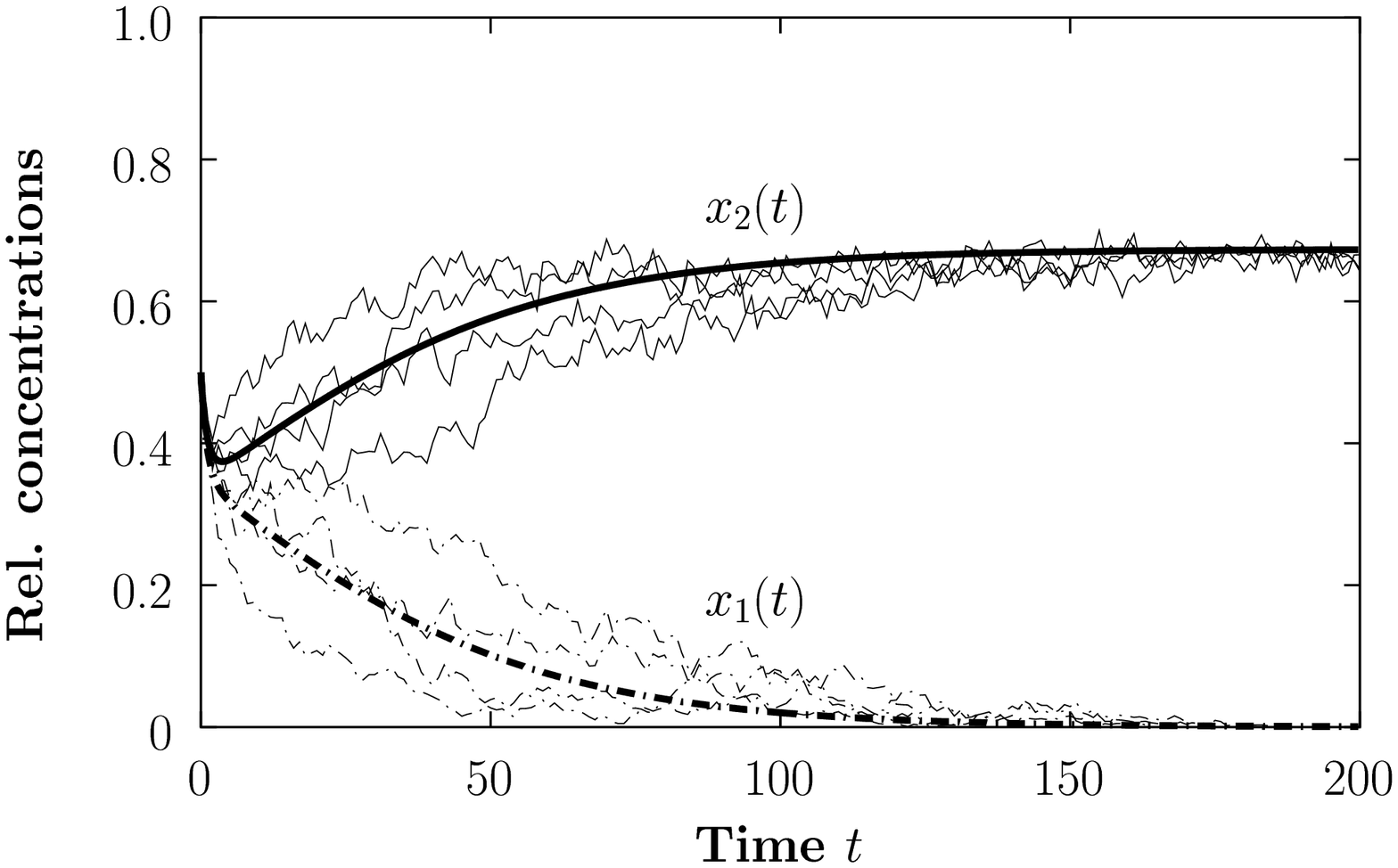}
}

\newpage
\noindent Figure 5:

\centerline{
\includegraphics[width=20cm,angle=90]{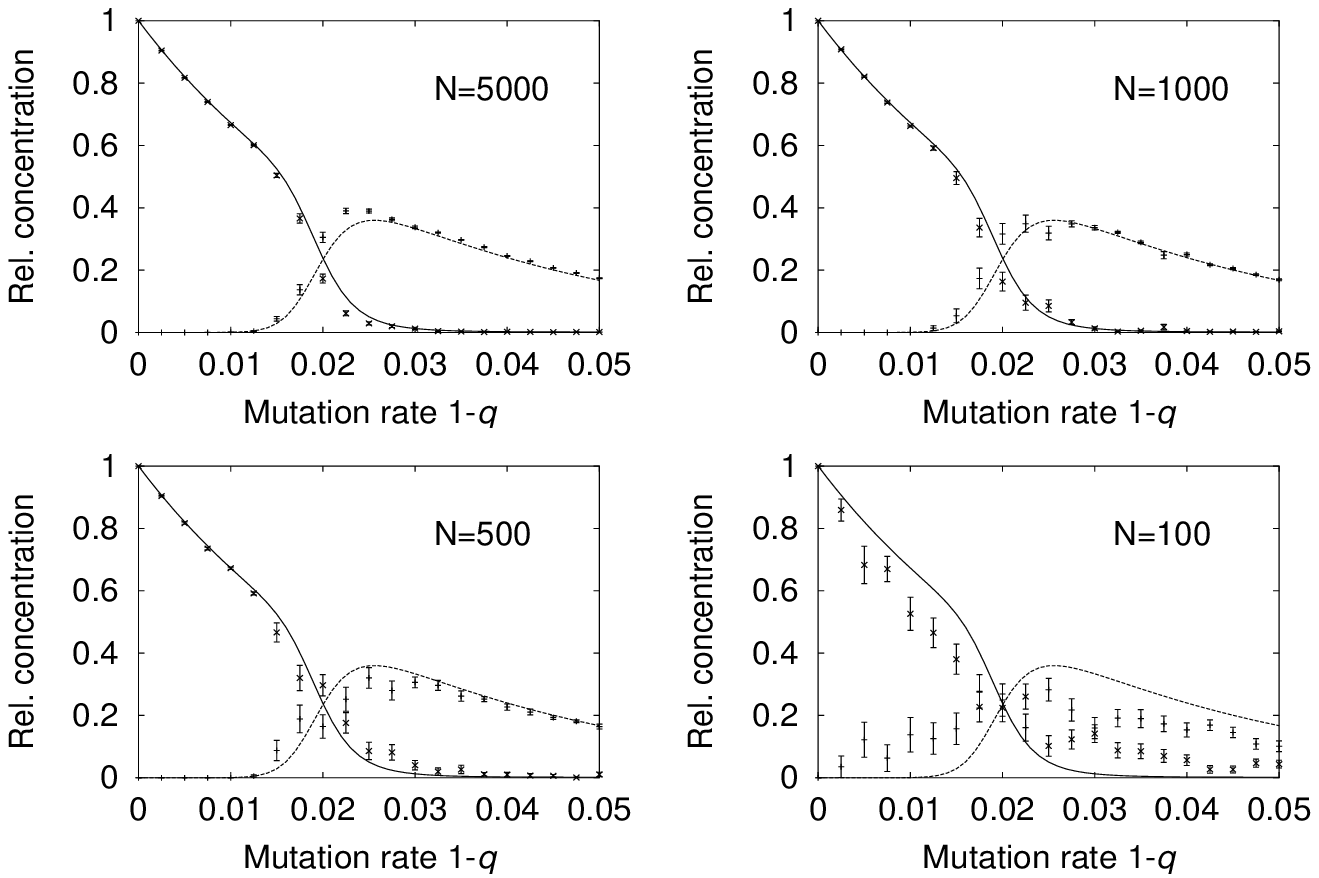}
}

\newpage
\noindent Figure 6:

\centerline{
\includegraphics[width=16cm,angle=90]{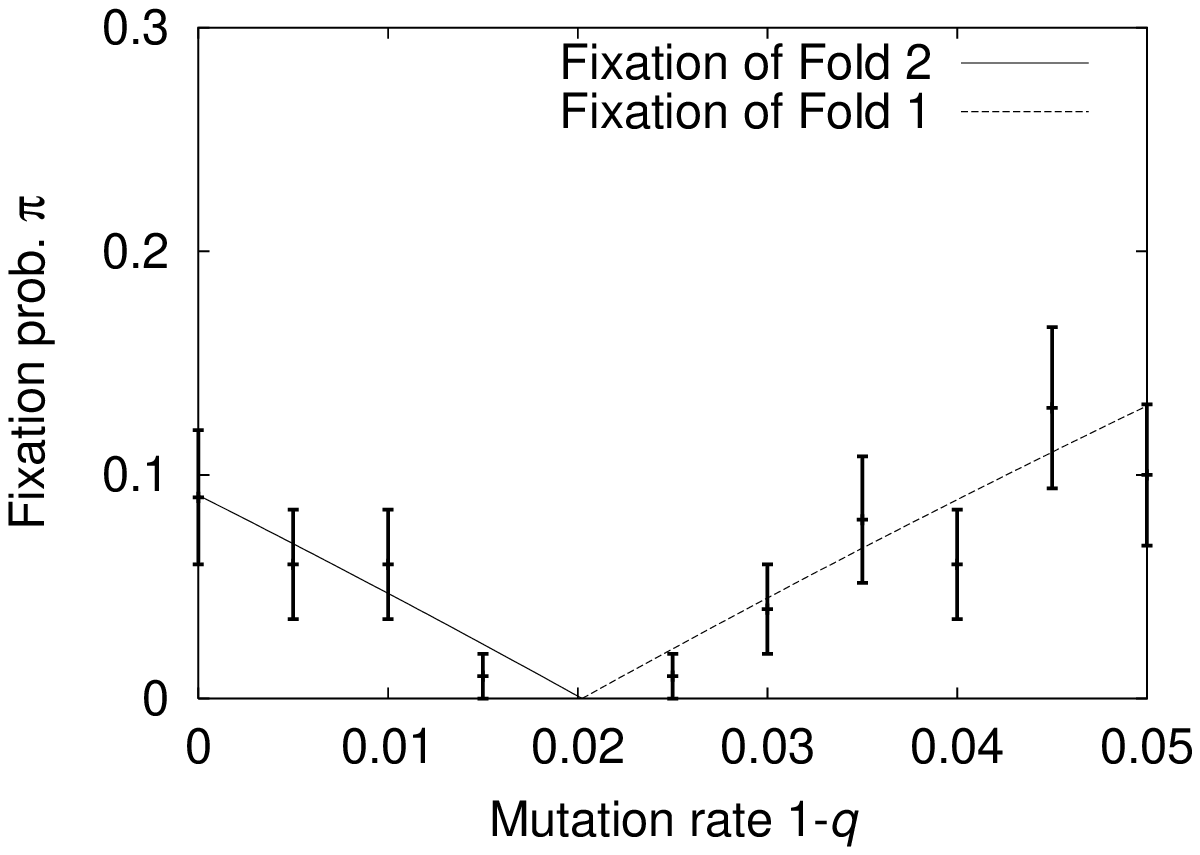}
}

\end{document}